\documentclass[10pt,conference]{IEEEtran}

\usepackage[utf8]{inputenc}
\usepackage[most]{tcolorbox}
\usepackage{xspace}
\usepackage{graphicx}
\usepackage{url}
\usepackage{booktabs}
\usepackage[flushleft]{threeparttable}
\usepackage{pifont}
\usepackage{colortbl}
\usepackage{multirow}
\usepackage{multicol}
\usepackage{makecell}
\usepackage{amssymb}
\usepackage{cite}
\usepackage{xcolor}
\usepackage{caption}

\newcommand{\app}[1]{\textit{#1}}
\newcommand{\code}[1]{\texttt{#1}}

\newcommand{\defectlib}{\textsc{Malva}\xspace}
\newcommand{\tool}{\textsc{MalvaGuard}\xspace}

\newcommand{\eg}{\hbox{\emph{e.g.}}\xspace}

\usepackage{pythonhighlight}

\definecolor{codegreen}{rgb}{0,0.6,0}
\definecolor{codegray}{rgb}{0.5,0.5,0.5}
\definecolor{codepurple}{rgb}{0.58,0,0.82}
\definecolor{backcolour}{rgb}{0.97,0.97,0.95}
\definecolor{forestgreen}{rgb}{0.28,0.62,0.37}

\lstdefinestyle{mystyle}{
    backgroundcolor=\color{backcolour},   
    commentstyle=\color{codegray},
    keywordstyle=\color{codepurple},
    numberstyle=\tiny\color{codegray},
    stringstyle=\color{blue},
    basicstyle=\ttfamily\footnotesize,
    breakatwhitespace=false,         
    breaklines=true,                 
    captionpos=b,                    
    keepspaces=true,                 
    numbers=left,                    
    numbersep=5pt,                  
    showspaces=false,                
    showstringspaces=false,
    showtabs=false,                  
    tabsize=4,
}

\usepackage{hyperref}

\IEEEoverridecommandlockouts

\begin{document}

\title{Understanding Tool-Mediated Interaction Failures\\ in Agentic GUI Software: An Empirical Study}

\author{%
\IEEEauthorblockN{Yuchen Shao\IEEEauthorrefmark{1}\IEEEauthorrefmark{2}, Wenjing Liu\IEEEauthorrefmark{1}, Yuheng Huang\IEEEauthorrefmark{3}, Mengqian Xu\IEEEauthorrefmark{1}\IEEEauthorrefmark{2}, Ziqun Bao\IEEEauthorrefmark{1}, Long Yang\IEEEauthorrefmark{1}\IEEEauthorrefmark{2}, \\[0.3ex]
Lei Ma\IEEEauthorrefmark{3}\IEEEauthorrefmark{4}, Ting Su\IEEEauthorrefmark{1}, Chengcheng Wan\IEEEauthorrefmark{1}\IEEEauthorrefmark{2}\thanks{Corresponding author: Chengcheng Wan (ccwan@sei.ecnu.edu.cn)}}
\IEEEauthorblockA{\IEEEauthorrefmark{1}East China Normal University, Shanghai, China\qquad
\IEEEauthorrefmark{2}Shanghai Innovation Institute, Shanghai, China\\[0.3ex]
\IEEEauthorrefmark{3}University of Tokyo, Tokyo, Japan\qquad
\IEEEauthorrefmark{4}University of Alberta, Edmonton, Canada}
}

\maketitle
\begin{abstract}
Large language model (LLM) agents increasingly complete tasks through live graphical user interfaces (GUIs). Because the model cannot directly operate the GUI, these systems often rely on tools to capture screenshots or UI trees, translate model outputs into GUI actions, and report whether the interface changed. We study failures where this tool-mediated interaction makes the agent's view of the task diverge from the live GUI. We construct \defectlib, a defect library with 335 confirmed real-world failures from 40 open-source applications. These failures fall into 11 defect patterns rooted in state representation, state transition, and synchronization problems. They are widespread: 90\% of the studied applications contain more than one defect type. They are also often silent: incorrectness appears in 86.6\% of failures, including 119 cases that produce wrong behavior without a fail-stop signal. We operationalize this taxonomy with \tool, a pre-deployment static contract checker that reaches 57.3\% precision and 76.8\% recall on an additional 20-application corpus.

\end{abstract}


\section{Introduction}
\label{sec:intro}
\sloppy

\subsection{Motivation}

With the rapid advancement of large language models (LLMs), an increasing number of software systems pair an LLM agent with tools that observe and operate existing graphical user interfaces (GUIs), including web browsers~\cite{webrover}, desktop applications~\cite{bytebot,openclaw2026}, and mobile applications~\cite{mobileagent}, without relying solely on programmatic APIs~\cite{zhou2024webarena,OSWorld,AppAgent}. This paper refers to these systems as \textit{agentic GUI software}~\cite{nguyen-etal-2025-gui,tang2025surveymllmbasedguiagents,jang2025videowebarena}. 
In the past 12 months, over 3,000 open-source applications have emerged on GitHub, along with around 30 commercial products launched by leading tech companies (\eg, Anthropic~\cite{anthropic}). 
At runtime, tools in these systems often capture GUI state from screenshots or UI trees, and the LLM plans the next step. Tools then execute actions such as clicks or typing and report the resulting environment feedback back to the agent~\cite{li2025surveyguiagentsfoundation}. 
While promising, this tool-mediated process introduces a reliability problem: the agent's view of the task can silently diverge from the live GUI~\cite{OSWorld}.

\begin{figure}
    \centering
    \includegraphics[width=\linewidth]{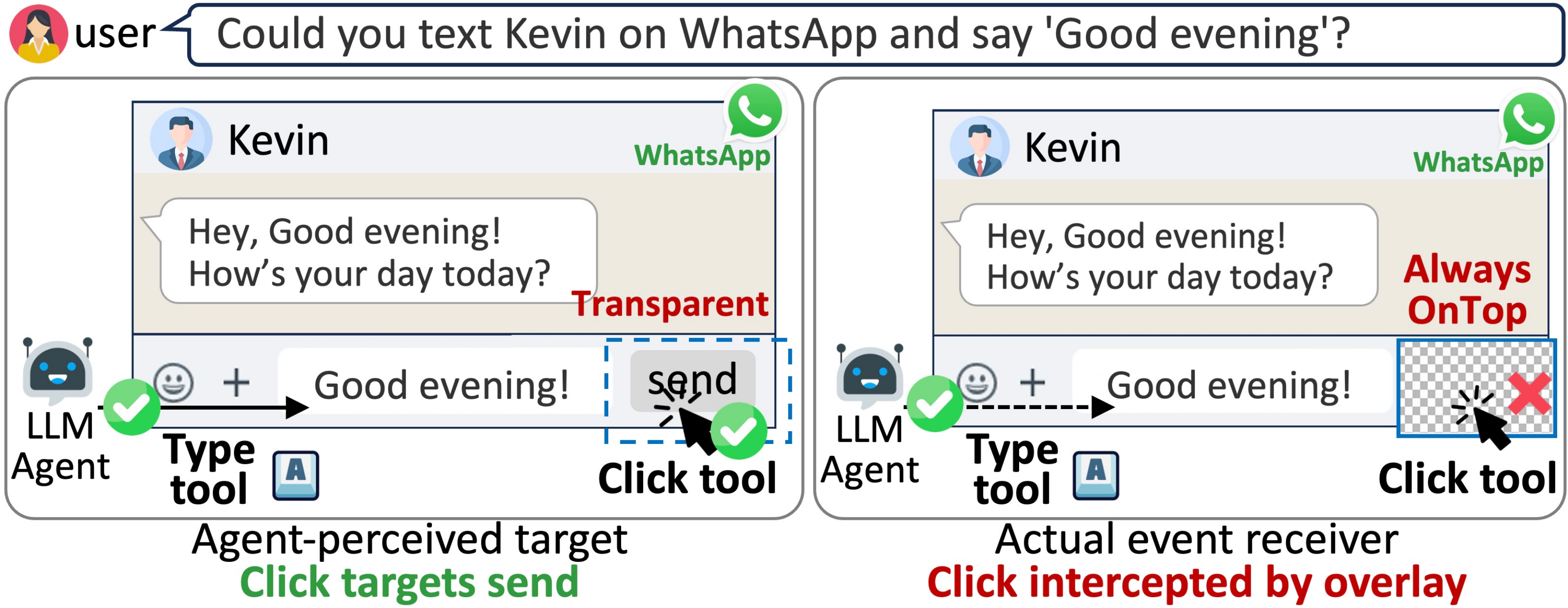}
    \vspace{-0.1in}
    \caption{An example of tool-mediated mismatch in \app{UI-TARS-desktop}~\cite{ui-tars-desktop}}.
    \label{fig:intro_example}
\end{figure}

Such divergence matters because these agents are moving into real digital workflows, including enterprise dashboards, customer-support portals, office productivity tools, online transactions, and mobile services~\cite{webrover,bytebot,anthropic}. In these settings, a tool misrouting a click can submit the wrong form, a stale screenshot can trigger redundant transactions, and premature success feedback can leave critical tasks silently incomplete~\cite{OSWorld}, resulting in financial loss, data corruption, or irreversible errors that are hard to detect and recover from.

To illustrate such defects, consider a real scenario where the vision-language desktop application \app{UI-TARS-desktop}~\cite{ui-tars-desktop} sends messages via WhatsApp~\cite{whatsapp}. As shown in Figure~\ref{fig:intro_example}, when a user instructs the agent to send a greeting, the agent minimizes its main interface and keeps a small transparent status window on top of the desktop. The \code{type} tool succeeds because the text field remains unobstructed. The next \code{click} tool, however, is issued for the send button at the perceived coordinates, while the OS routes the event to the agent's own overlay instead, and the message is never sent. The failure exposes a missing check: the tool exposes visual coordinates without verifying which window holds focus. This failure is not isolated, developers of agentic GUI software face three reliability challenges:

\begin{figure*}
    \centering
    \includegraphics[width=0.83\linewidth]{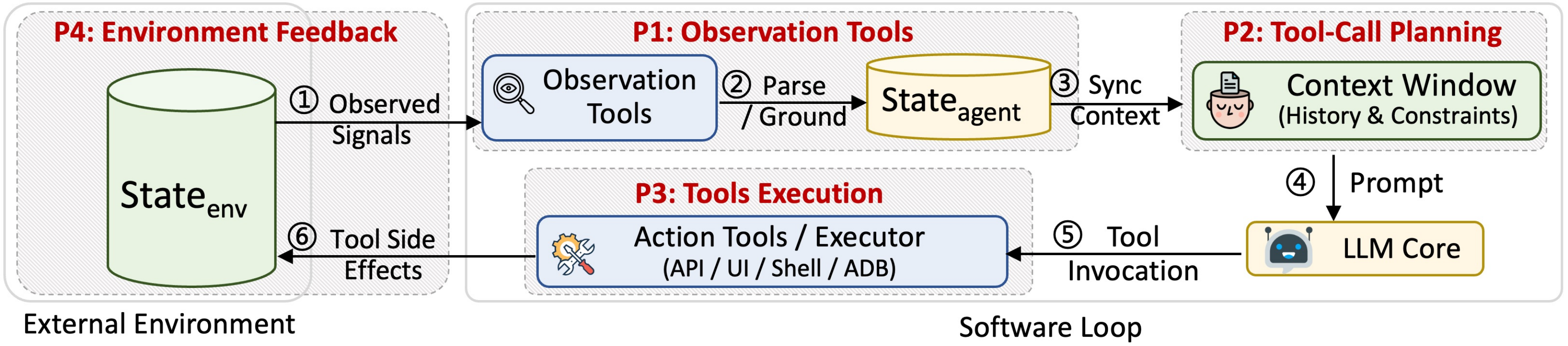}
    \vspace{-0in}
    \caption{Tool-mediated workflow of agentic GUI software. }
    \label{fig:interaction-workflow}
\end{figure*}

\textbf{Challenge-1: Partial GUI observations.} Developers often need agents to operate applications without programmable interfaces, relying instead on tools that read screenshots or UI trees~\cite{whatwg-dom-standard,nguyen-etal-2025-gui}. These tools expose only a slice of the live interface: elements may be obscured, off-screen, omitted from the UI tree, or visually present but not actionable. As a result, the agent may plan from an incomplete view of the target state.

\textbf{Challenge-2: Weak action boundaries.} 
Developers expose large action spaces so agents can handle diverse tasks, but they rarely specify each tool's operational boundary, including accepted parameters, required GUI preconditions, expected return format, or when to stop, retry, or escalate~\cite{patil2024gorilla}. This leaves the tool loop unable to determine whether the next step is valid, leading to malformed calls, non-progressing retries, context overflow, or actions that target the wrong element.

\textbf{Challenge-3: Delayed tool feedback.} 
LLM inference, GUI rendering, and tool execution run at different speeds~\cite{10.1145/2635868.2635920}. After an action is executed, transient GUI changes, backend timeouts, or expired resource handles can cause reported tool status to diverge from the actual live interface state before the next observation begins.

Despite extensive work on GUI automation and LLM agents, the tools that connect agents to live GUI environments have received little attention. Empirical work studied defects in traditional software~\cite{large-scale_study,DeepBugs}, ML-enabled systems~\cite{10.1145/3377811.3380395,ML_API,keeper,9000651}, and LLM integrations~\cite{llm_study,zhang2025which}. Traditional GUI automation treats automation code as test infrastructure, through scripted, model-based, and AI-assisted approaches~\cite{selenium2026,guitar,8952324}, while GUI-agent studies evaluate task completion across web~\cite{zhou2024webarena,deng2023mindweb}, OS~\cite{OSWorld}, and mobile~\cite{rawles2023androidinthewild} settings, and frameworks such as OpenClaw~\cite{openclaw2026} propose architectural designs. Across all these lines, prior work either treats the tool layer as invisible infrastructure or focuses on model and agent behavior, overlooking the tools that mediate between the agent and the environment. This paper fills this gap by systematically studying how such tools introduce failures in agentic GUI software.

\subsection{Contribution}
To understand how tools introduce failures when LLM agents operate live GUI environments, we conduct a systematic study of 40 GitHub projects, based on their latest versions as of \textit{January 1\textsuperscript{st}, 2026}. These applications enable LLM agents to autonomously operate graphical interfaces across web, desktop, and mobile platforms. We manually collect around 2,000 issue reports and conduct extensive task execution to confirm 335 such failures.

We find that such failures are widespread: 90\% of the studied applications contain multiple defect types. The failures cause incorrect task results, infinite execution loops, wasted tokens, and unexpected interruptions. As illustrated in Figure~\ref{fig:interaction-workflow}, they occur across four phases of the software loop: tools first observe and ground GUI state, the agent assembles context and selects the next tool call, tools then perform physical or API-backed actions, and tools finally report execution status and environment changes back to the agent.

We identify 11 common defect patterns caused by state representation, state transition, and synchronization problems. We release \defectlib, a defect library containing all 335 confirmed failures with their root causes and impacts. Based on the findings, we build \tool, a pre-deployment static contract checker for such failures in agentic GUI software. On an additional 20-application corpus, it achieves 57.3\% precision and 76.8\% recall. We hope this work can serve as a starting point for improving the reliability of agentic GUI software.

\section{Background}

\label{sec:bg}

\sloppy

\begin{figure*}
    \centering
    \includegraphics[width=0.9\linewidth]{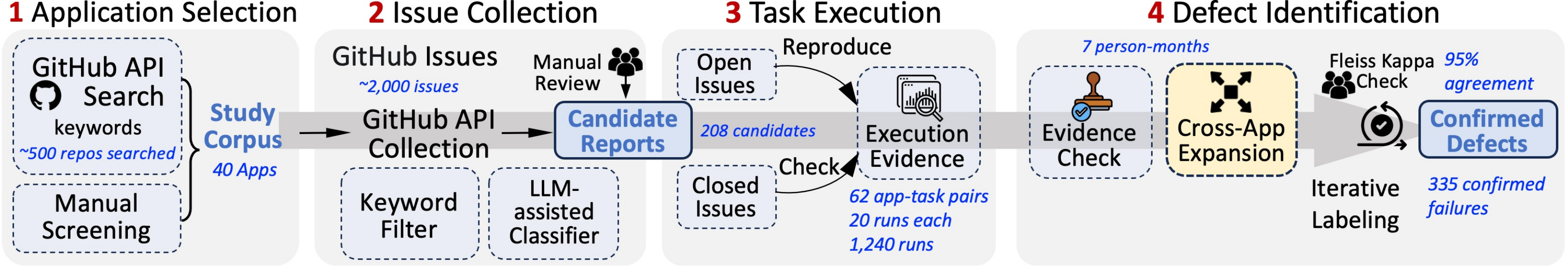}
    \vspace{-0in}
    \caption{Overview for constructing \defectlib defect library.}
    \label{fig:method_pipeline}
\end{figure*}

\subsection{Agentic GUI Software}
Agentic GUI software lets an LLM agent operate a live web, desktop, or mobile 
environment through a tool layer~\cite{gur2024a,10.5555/3600270.3601778,OSWorld,AppAgent}. 
Although this paper studies the full system, we refer to these \textit{applications} as \textit{agents} 
throughout for simplicity. 
The agent maintains $\text{State}_{\text{agent}}$, which consists of the context window sent to the LLM together with memory, observations, action history, reasoning outputs, and selected tool calls. The environment has its own $\text{State}_{\text{env}}$, covering the current GUI widgets, windows, focus, permissions, loading state, network conditions, and received tool-side effects.

The \textit{tool layer} is the bridge between the agent and the environment, and its tools fall into three roles. \textit{Observation tools} collect GUI evidence for the agent, such as screenshots, UI trees, Document Object Model (DOM)~\cite{whatwg-dom-standard} snapshots, OCR text, and element coordinates. Based on this evidence, the agent selects an action, which \textit{action tools} then translate into backend effects such as clicks, typing, navigation, shell commands, browser API calls, mobile gestures, or ADB~\cite{adb} commands, sometimes inside sandboxed browsers, desktops, or virtual machines~\cite{zhou2024webarena,OSWorld}. \textit{Feedback tools} then close the loop by reporting execution status and visible effects back to the agent, including success signals, errors, timeouts, page loads, and target changes.

\subsection{Tool-Mediated Interaction Loop}
Given these components, agentic GUI software runs as a tool-mediated feedback loop~\cite{yao2023react,GPT-4V}. Formally, one cycle is defined as
$$(\text{State}_{\text{env}, t}, \text{State}_{\text{agent}, t}) \xrightarrow{\mathcal{T}_t} (\text{State}_{\text{env}, t+1}, \text{State}_{\text{agent}, t+1})$$
where $\mathcal{T}_t$ denotes the tool-mediated transition across observation, invocation, execution, and feedback. During this transition, tools may change the live GUI, append observations or action results to the agent state, and update tool-local state such as caches, parsed snapshots, and backend handles. A later step therefore depends not only on the LLM output, but also on whether these tool states remain aligned with the live environment.

Figure~\ref{fig:interaction-workflow} splits the loop into four phases: P1 observes and grounds GUI state; P2 places relevant agent state into the context window and asks the LLM for the next action; P3 parses that output and executes the tool or backend operation; and P4 reports effects back to the agent. Reliable interaction therefore requires that observations reflect the current GUI, action parameters match valid targets, the GUI is ready before execution, and feedback confirms real progress.

\subsection{Interaction Failures}
When tools fail to keep $\text{State}_{\text{agent}}$ aligned with $\text{State}_{\text{env}}$, the agent operates on a false view of the task. We define this mismatch as a \textit{tool-mediated interaction failure}, or \textit{interaction failure} for short. It covers hard failures such as crashes and exceptions, and soft failures where the loop stalls, repeats unproductive actions, or completes the wrong task~\cite{AgentBoard,AgentBench}. A defect occurs when a tool accepts or executes an invalid state without the necessary validation or effect check.

\section{Study Methodology}
\label{sec:methodology}
Since no prior work has systematically studied failures at the tool-mediated boundary in agentic GUI software, we construct \defectlib, a defect library of confirmed failures, from scratch. 
As summarized in Figure~\ref{fig:method_pipeline}, we follow four procedures, each described in detail below:

\subsection{Application Selection}

We select 40 open-source agentic GUI applications from GitHub, using their latest versions as of \textit{January 1\textsuperscript{st}, 2026}. We first use the GitHub Search API~\cite{github_api} to collect about 500 repositories of agentic GUI software using keywords that cover major platforms and agent types, such as ``GUI agent'', ``web agent'', ``desktop agent'', ``mobile agent'', and ``computer use agent''.  
After removing duplicates and inactive forks, we keep repositories that meet three criteria:
\begin{enumerate}
    \item the repository implements an executable agent loop for real-world tasks, not a toy demo or thin API wrapper;
    \item the implementation contains concrete observation and action tools for GUI interaction;
    \item the repository contains enough implementation and issue evidence to support defect analysis.
\end{enumerate}
We further review around 200 repositories from the same GitHub search space, but find no major differences from the existing corpus. We therefore stop the collection.

The applications target three environments, as shown in \autoref{tbl:statistics}: 40\% are web agents that browse and complete online tasks, 43\% are desktop agents that use mouse and keyboard simulation to operate software and manage files, and 18\% are mobile agents that use swipes and taps to operate phone applications. Most are written in Python (68\%) or TypeScript (25\%), with the rest using other languages. For reasoning, 60\% use cloud APIs, mainly GPT~\cite{openai2024}, 8\% use local models, mainly LLaMA~\cite{llama_meta}, and 33\% support both. For GUI observation, 58\% rely only on screenshots, 15\% rely only on structural UI representations such as DOM trees, and 27\% combine both. Their primary grounding mechanisms also vary: 38\% use direct screenshot-to-coordinate VLM grounding~\cite{wu2025osatlas}, 15\% add detectors, OCR, or grounding modules, 23\% rely on accessibility or UI-tree structures, and 25\% use DOM/XPath, browser-structural, or task-specific grounding. 

The applications average 71,687 lines of code, with a median of 24,978 and an interquartile range from 2,910 to 57,146. They have 226 commits on average, a median of 438 stars, and a maximum of 19,330 stars. In addition, 63\% are less than one year old, reflecting how quickly the field is changing. Among desktop and mobile agents, 96\% support workflows across multiple applications, while 4\% focus on a single application.

\begin{table}
\footnotesize
\centering
\caption{Statistics of studied applications}
\label{tbl:statistics}
\vspace{-0.05in}
\resizebox{\linewidth}{!}{
    \begin{tabular}{l|ccccc}
    \hline
    \textbf{Environment}    & \textbf{\# of Projects} & \textbf{LOC} & \textbf{Stars} & \textbf{Commits} & \textbf{Age (Mo.)} \\ \hline
    Web agents &     16       &  48162  &   1402    &       130 &       14 \\
    Desktop agents  &     17       &  89643  &   3395    &       366 &       13 \\
    Mobile agents  &     7    &  81849  &   2016    &       109 &       10 \\ \hline
    \end{tabular}
}
\begin{flushleft}
\scriptsize \textit{* Values in the last four columns represent the average across all projects in each environment.}
\end{flushleft}

\label{tab:dataofapps}
\end{table}

\subsection{Issue Collection}

We use the GitHub API to collect all non-empty issue reports with active links from the 40 applications, yielding about 2,000 unique reports. We exclude reports tagged as dependency or infrastructure issues, or those attributed to ambiguous user prompts. We then apply keyword filters and an LLM-assisted classifier~\cite{3666122.3668142} with GPT-4o~\cite{openai_gpt4o} for a preliminary screening. Our team subsequently reviews the filtered results manually and retains 208 candidates for confirmation.

\begin{table*}
  \centering
  \caption{Defects identified in agentic GUI software.}
  \label{tbl:identified_defects}
  \vspace{-0.08in}
  \scriptsize
  \setlength{\tabcolsep}{3.5pt}
  \resizebox{\textwidth}{!}{%
      \begin{threeparttable}
          \begin{tabular}{lccccccccccccc}
          \Xhline{1pt}
          \multirow{2}{*}{\textbf{Defect}} &
            \multirow{2}{*}{\textbf{Section}} &
            \makecell[c]{\textbf{Violated}} &
            \makecell[c]{\textbf{Root}} &
            \multicolumn{6}{c}{\makecell[c]{\textbf{Impact}}} &
            \makecell[c]{\textbf{Problematic}} &
            \multicolumn{3}{c}{\makecell[c]{\textbf{Environment}}} \\ \cline{5-10} \cline{12-14}
            & 
              & 
            \textbf{Tool Contract} &
            \textbf{Cause} &
            ST &
            IC &
            SL &
            UI &
            TK &
            IS &
            \textbf{Apps (\%)} &
            Web &
            Desktop &
            Mobile \\ \Xhline{1pt}
          \rowcolor{lightgray!50} \multicolumn{14}{c}{\textbf{Phase 1: Observation Defects}} \\ \hline
          Observation Hallucination &
            \hyperref[subsubsec:UI_hallucination]{\S\ref{subsec:P1_perception}.1} &
            Observation &
            SRF &
            \checkmark &
            \checkmark &
             &
            \checkmark &
            \checkmark &
             &
            33\% &
            4 &
            8 &
            1 \\ \hline
          Target Grounding Error &
            \hyperref[subsubsec:element_misidentification]{\S\ref{subsec:P1_perception}.2} &
            Target grounding &
            SRF &
            \checkmark &
            \checkmark &
            \checkmark &
             &
             &
             &
            38\% &
            5 &
            8 &
            2 \\ \hline
          State Exposure Gap &
            \hyperref[subsubsec:missing_state_awareness]{\S\ref{subsec:P1_perception}.3} &
            State exposure &
            SRF &
            \checkmark &
            \checkmark &
             &
            \checkmark &
             &
             &
            18\% &
            3 &
            2 &
            2 \\ \hline
          \rowcolor{lightgray!50} \multicolumn{14}{c}{\textbf{Phase 2: Reasoning \& Planning Defects}} \\ \hline
          Progress Check Failure &
            \hyperref[subsubsec:infinite_loop]{\S\ref{subsec:P2_reasoning}.4} &
            Progress &
            STF &
            \checkmark &
            \checkmark &
            \checkmark &
             &
            \checkmark &
             &
            65\% &
            11 &
            12 &
            3 \\ \hline
          Escalation Boundary Failure &
            \hyperref[subsubsec:permission_security]{\S\ref{subsec:P2_reasoning}.5} &
            Escalation &
            STF &
            \checkmark &
             &
             &
            \checkmark &
            \checkmark &
            \checkmark &
            5\% &
            2 &
            0 &
            0 \\ \hline
          Context Budget Overflow &
            \hyperref[subsubsec:token_limit_exceeded]{\S\ref{subsec:P2_reasoning}.6} &
            Context budget &
            STF &
            \checkmark &
             &
             &
             &
            \checkmark &
             &
            48\% &
            6 &
            13 &
            0 \\ \hline
          \rowcolor{lightgray!50} \multicolumn{14}{c}{\textbf{Phase 3: Execution Defects}} \\ \hline
          Tool Schema Mismatch &
            \hyperref[subsubsec:format_schema_mismatch]{\S\ref{subsec:P3_execution}.7} &
            Schema &
            STF &
            \checkmark &
            \checkmark &
             &
             &
            \checkmark &
             &
            63\% &
            7 &
            13 &
            5 \\ \hline
          Parameter Semantics Error &
            \hyperref[subsubsec:parameter_hallucination]{\S\ref{subsec:P3_execution}.8} &
            Parameter semantics &
            STF &
            \checkmark &
            \checkmark &
             &
             &
             &
             &
            25\% &
            6 &
            3 &
            1 \\ \hline
          Missing Execution Preconditions &
            \hyperref[subsubsec:precondition_violation]{\S\ref{subsec:P3_execution}.9} &
            Preconditions &
            STF &
            \checkmark &
            \checkmark &
             &
             &
             &
             &
            33\% &
            4 &
            7 &
            2 \\ \hline
          \rowcolor{lightgray!50} \multicolumn{14}{c}{\textbf{Phase 4: Environment Feedback Defects}} \\ \hline
          Temporal Feedback Mismatch &
            \hyperref[subsubsec:latency_mismatch]{\S\ref{subsec:P4_environment}.10} &
            Temporal feedback &
            SF &
            \checkmark &
            \checkmark &
            \checkmark &
             &
            \checkmark &
             &
            20\% &
            3 &
            5 &
            0 \\ \hline
          Lifecycle Propagation Failure &
            \hyperref[subsubsec:unexpected_interruption]{\S\ref{subsec:P4_environment}.11} &
            Lifecycle/error &
            SF &
            \checkmark &
            \checkmark &
             &
            \checkmark &
             &
             &
            58\% &
            8 &
            14 &
            1 \\ \hline
          \Xhline{1pt}
          \rowcolor{lightgray!20}
          \multicolumn{10}{r}{Total applications with \emph{at least 2} defect types} & \textbf{90\%} & \multicolumn{3}{c}{} \\ \Xhline{1pt}
          \end{tabular}
          \begin{tablenotes}
              \scriptsize
              \item * Environment columns: Web/Desktop/Mobile numbers count affected applications.
              \item * Impact abbreviations: ST=fail-stop, IC=incorrectness, SL=slower execution, UI=user-interface disruption, TK=increased token usage, IS=security exposure.
          \end{tablenotes}
      \end{threeparttable}%
  } 
  \vspace{-0.08in}
\end{table*}

\subsection{Task Execution}
To collect evidence for candidate reports, we first inspect the issue status: 104 are closed issues, and 104 are open issues.
\begin{itemize}
    \item \textbf{Closed issues}: we examine discussions, linked pull requests, and commits to identify the failure scenario that developers acknowledged or fixed.
    \item \textbf{Open issues}: we reproduce the failure directly when the required environment and credentials are available.
\end{itemize}

Through this process, we derive 62 executable tasks from the open issue scenarios, since multiple issues often describe the same scenario. For each open issue, we identify the user goal, required environment state, and expected outcome. Each task specifies an initial state, an agent instruction, and a success oracle such as a UI change or system confirmation, instantiated under 14 environment-specific scenarios. The tasks span web, desktop, and mobile agents, with 35, 14, and 13 tasks respectively, and range from simple actions such as clicking or text entry to multi-step planning such as booking flights or configuring a system.

To account for non-determinism, we run each of the 62 tasks 20 times with a 3-minute timeout per run~\cite{zhou2024webarena}, yielding 1,240 task runs. Each run involves multiple agent-environment interactions, and automated logging records visual evidence including screen recordings and DOM snapshots, agent traces including prompts and action logs, tool invocations and return values, and environment feedback such as page loads and errors. In total, the dataset contains about 3,000 agent interactions.

We conduct most experiments on Windows 11 to ensure a consistent graphical environment, using Linux virtual machines for Linux-only applications.

\subsection{Defect Identification}
Using the evidence collected in the previous procedure, we manually assess each of the 208 candidates to verify whether the reported failure is real and reproducible:
\begin{itemize}
    \item \textbf{Confirmed} when multiple evidence sources converge on the same tool-layer mistake, such as linked fixes or commits for closed issues, and reproduction logs, tool traces, and before/after GUI states for open issues.
    \item \textbf{Excluded} when the failure only appears in a few runs due to unstable environments, missing credentials, or unsupported setups.
\end{itemize}
This yields 99 confirmed defects out of the 208 candidates, including 43 from open issues and 56 from closed issues.

As a failure confirmed from one GitHub issue may not be unique to that application, we next perform cross-app expansion. For each of the 99 confirmed tool-layer mistakes, we exhaustively inspect all remaining 39 applications to check whether the same erroneous tool behavior arises at the same type of tool-environment interaction. When a matching case is found, we reproduce the corresponding failure and count each successful reproduction as a separate confirmed failure. This process yields 335 confirmed interaction failures in total.

Finally, we label the confirmed failures through a separate iterative process. In each iteration, four authors discuss the identified failures and refine a list of patterns based on workflow phase, root cause, tool-layer mistake, and impact. Three authors independently categorize each failure into one of the defect patterns, label its impact, and cross-check their labels, reaching 95\% initial agreement (Fleiss' $\kappa=0.95$). A fourth author joins when disagreements remain and resolves them by revisiting the original evidence. The iterations stop when no new patterns emerge, and the labeling effort took about 7 person-months.

\section{Defect Pattern Analysis}
\label{sec:defect}

\sloppy
\subsection{Overview} 

We classify 335 defects across 40 GitHub applications into 11 defect patterns, summarized in \autoref{tbl:identified_defects}. The taxonomy uses two axes to capture both where in the loop the failure occurs and what tool-layer contract is violated. The phase axis locates the earliest broken tool-layer boundary across P1 to P4. The root-cause axis records which contract is violated between the agent state, tool behavior, and live GUI state:
\begin{itemize}
    \item \textbf{State Representation Failure (SRF, 26.3\%)}: Observation or grounding tools build $\text{State}_{\text{agent}}$ from GUI evidence that is missing, wrong, stale, or incomplete, corresponding to Challenge-1.
    \item \textbf{State Transition Failure (STF, 55.5\%)}: The tool loop or executor advances the interaction even though the next step is invalid, unsupported, not ready, not progressing, or over budget, reflecting Challenge-2.
    \item \textbf{Synchronization Failure (SF, 18.2\%)}: Tool feedback or lifecycle management code updates $\text{State}_{\text{agent}}$ with a status that no longer matches $\text{State}_{\text{env}}$ or available resources, mirroring Challenge-3.
\end{itemize}

The failures concentrate around recurring tool-layer contracts instead of isolated GUI glitches. As shown in \autoref{tbl:identified_defects}, Progress Check Failure, Tool Schema Mismatch, and Lifecycle Propagation Failure each appear in more than half of the studied applications at 65\%, 63\%, and 58\% respectively, confirming that these defects are systemic weaknesses instead of application-specific bugs. At the failure level, these three patterns together with Context Budget Overflow account for 63.0\% of confirmed failures, with P2 alone contributing 31.0\% where the loop accepts the next step before checking progress, budget, or escalation conditions.

\begin{figure}
    \centering
    \includegraphics[width=\linewidth]{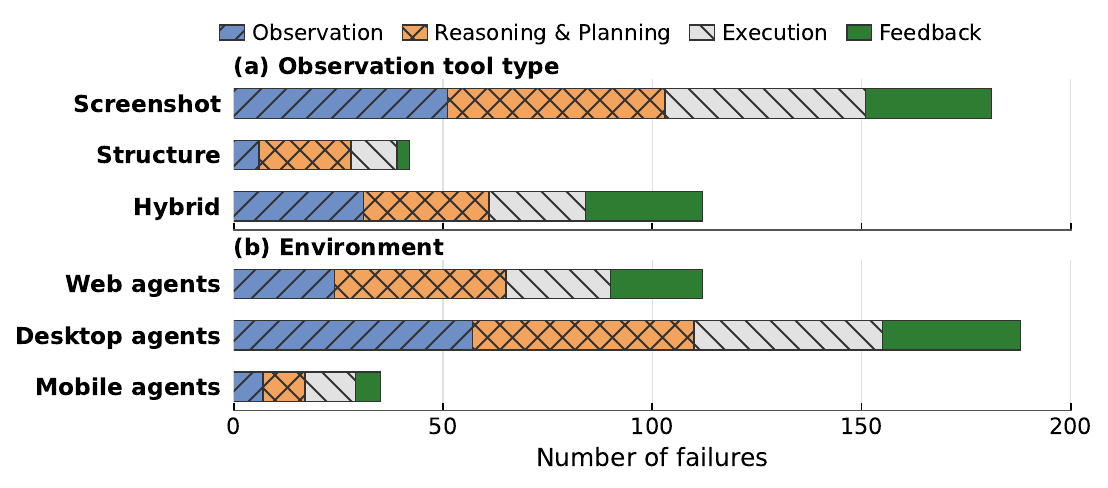}
    \vspace{-0.1in}
    \caption{Defect distribution across workflow phases by (a) observation tool type and (b) environment.}
    \label{fig:tool_layer_phase_distribution}
\end{figure}

This distribution further varies by observation tool type and execution environment, as shown in Figure~\ref{fig:tool_layer_phase_distribution}. Agents that rely solely on screenshots or structural snapshots concentrate failures in the phases tied to their grounding strategy: screenshot-based agents fail more in observation and execution, while structure-only agents fail more in P2 where DOM snapshots are mistaken for progress evidence. Desktop agents account for 56.1\% of all confirmed failures and dominate in execution and resource-lifetime patterns, whereas mobile and web agents concentrate in P3 and P2 respectively, reflecting their distinct action interfaces and loop control patterns.

The common mistake across all settings is treating model-visible state as if it already matched the live GUI: observations hide or perturb state, feedback tools report success before GUI changes, and tools promise more than the underlying execution layer can deliver. As a result, incorrectness appears in 86.6\% of confirmed failures, including 119 that produce wrong behavior without a fail-stop signal.

\subsection{Phase 1: Observation Defects}
\label{subsec:P1_perception}
P1 builds $\text{State}_{\text{agent}}$ from GUI evidence captured or carried by tools, including screenshots, structural snapshots, detected elements, and grounding results. When this evidence does not reflect the live GUI, the agent plans from a representation that no longer matches $\text{State}_{\text{env}}$, producing 88 of 335 failures. Across all three patterns, the common mistake is that tools expose GUI evidence without verifying it against the same live target the executor will act on.

\begin{figure}
    \centering
    \includegraphics[width=0.95\linewidth]{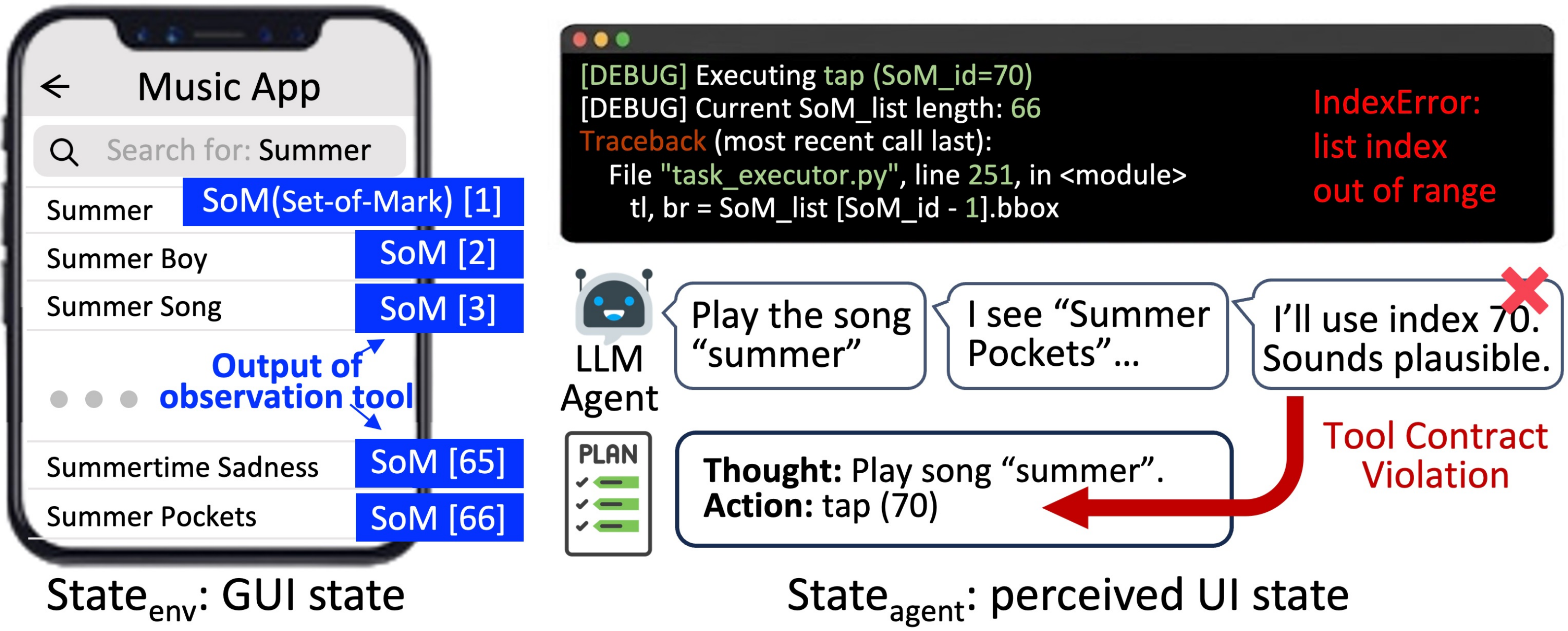}
    \vspace{-0in}
    \caption{Observation hallucination with phantom index in \app{AppAgent}~\cite{AppAgent}.}
    \label{fig:UI_Hallucination}
\end{figure}

\medskip
\phantomsection
\label{subsubsec:UI_hallucination}
\noindent\textbf{1. Observation Hallucination.}
Observation, completion, and indexed-action tools rely on existence checks before they update $\text{State}_{\text{agent}}$ or invoke an action. Without such checks, the agent may report fake completion or issue actions against nonexistent targets. We observe this pattern in 13 applications (33\%), averaging 2.69 cases each.

Hallucinations appear in two forms. First, \emph{fake-success} cases accept task completion without state evidence. The web automation application \app{gemini-browser}~\cite{geminibrowser2024} declares a file-saving task done without writing the file, and a DOM-aware browser case reports a UI switch although the state is unchanged~\cite{sinecan2024agentic_ai_browser}. Second, \emph{phantom-target} cases serialize absent evidence into actions. Figure~\ref{fig:UI_Hallucination} shows the multimodal smartphone GUI automation agent \app{AppAgent}~\cite{AppAgent} issuing \code{tap(70)} when only 66 detected elements exist, causing an \code{IndexError}. Both forms lack an existence check before $\text{State}_{\text{agent}}$ is updated or an action is invoked.

\medskip
\phantomsection
\label{subsubsec:element_misidentification}
\noindent\textbf{2. Target Grounding Error.}
Grounding and targeting tools bind an intended target to the element, coordinate, box ID, or controllable region that the executor can safely operate. A wrong binding turns a valid target into an invalid or wrong locator before the action tool runs. The pattern appears in 38\% of applications, averaging 2.60 cases each.

\begin{figure}
    \centering
    \includegraphics[width=0.8\linewidth]{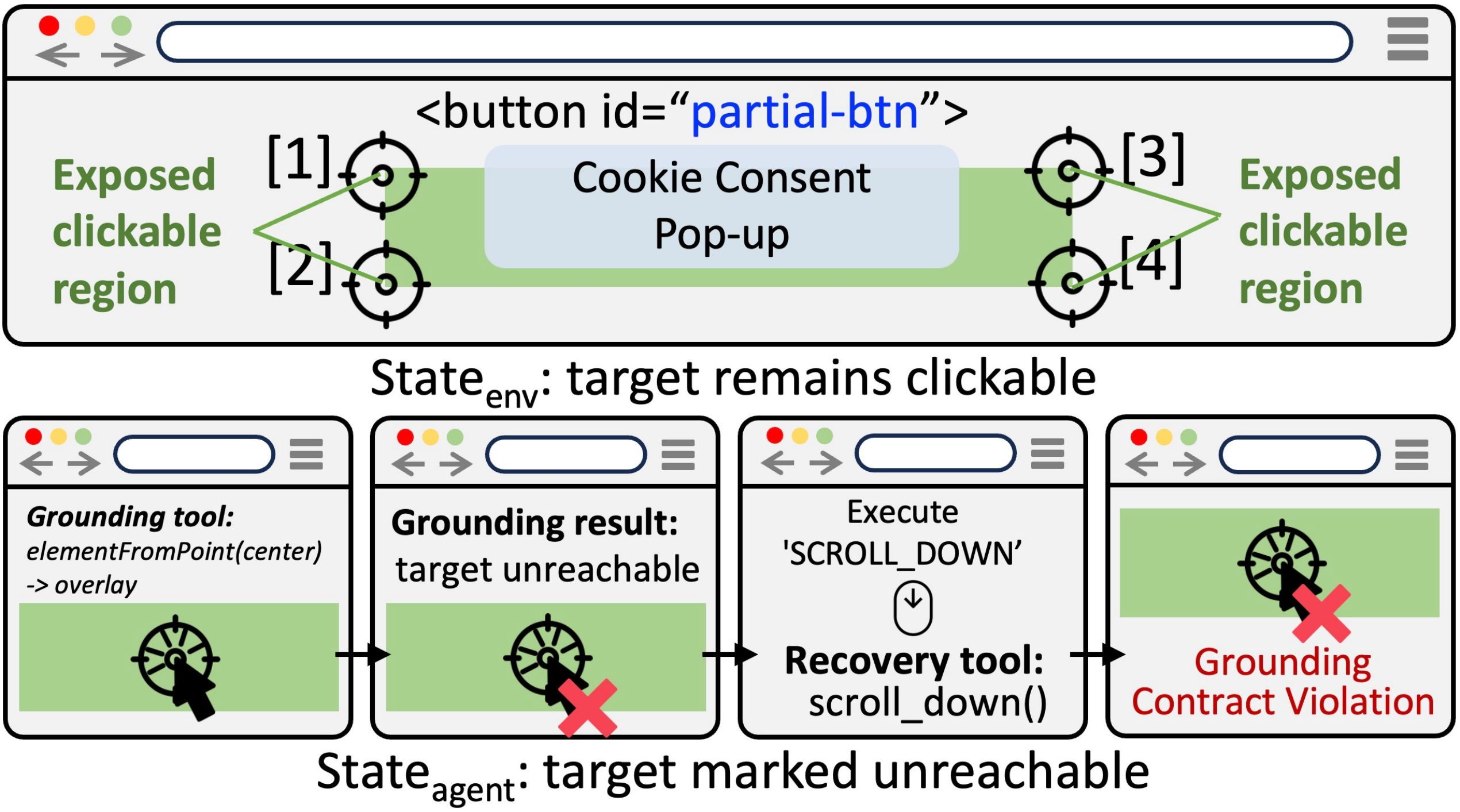}
    \vspace{0in}
    \caption{Center-point grounding failure in \app{LaVague}~\cite{lavague2026}.}
    \label{fig:element_misidentification}
\end{figure}

The mismatch stems from rigid geometry rejections, coordinate drift between tool components, or semantic confusion over UI context. Figure~\ref{fig:element_misidentification} shows the web automation agent \app{LaVague}~\cite{lavague2026} rejecting a widget because its center point is covered, even though other clickable regions remain exposed; the grounding tool therefore turns a valid target into an invalid locator. Coordinate conventions can drift between tool components, as in the multimodal desktop agent \app{UI-TARS-desktop}~\cite{ui-tars-desktop}, where JavaScript \code{Math.round} and Python banker's rounding disagree before a coordinate is sent to the executor. Semantic confusion appears when the grounding tool maps a requested target to the wrong UI context, such as the out-of-the-box desktop GUI agent \app{computer\_use\_ootb}~\cite{computer_use_ootb2026} confusing taskbar icons with browser launchers.

\medskip
\phantomsection
\label{subsubsec:missing_state_awareness}
\noindent\textbf{3. State Exposure Gap.}
Observation and grounding tools need to expose the live GUI state that can affect the next action, not only the visible or currently serialized slice. Missing obstruction, focus ownership, scroll reachability, modal state, or freshness leaves $\text{State}_{\text{agent}}$ with only a partial view of $\text{State}_{\text{env}}$. This pattern affects 18\% of applications, with a median of 2 cases each.

These gaps appear at several tool boundaries. At the window and focus boundary, as also seen in \app{UI-TARS-desktop}~\cite{ui-tars-desktop}, the screenshot tool exposes the target visually but not the topmost click owner, so the action is delivered to the wrong window layer, as illustrated in Figure~\ref{fig:intro_example}. The desktop automation agent \app{Clevrr-Computer}~\cite{clevrr-computer} exposes a related gap: its \code{get\_screen\_info()} tool always reports \code{1920 x 1080} to the vision model regardless of the actual display size, leaving the agent grounding targets against a stale window model instead of the active screen. At the viewport and scroll boundary, observation tools expose only the visible slice: the GPT-4V grounded web agent \app{SeeAct}~\cite{seeact2026} misses that a table column named ``topics'' is reachable through a horizontal scroll container, repeatedly concludes that no candidate matches the target, and stops. At the modal and loading boundary, observation tools keep reporting the underlying page while an unmodeled dialog or transient state hides the actionable UI.

\begin{tcolorbox}[colback=black!5!white,colframe=black!80!white,boxsep=1pt,left=2pt,right=2pt,top=1pt,bottom=1pt,arc=0pt,outer arc=0pt,boxrule=0.7pt]
\emph{Finding-1:} P1 accounts for 26.3\% of failures: observation and grounding tools expose GUI evidence that is nonexistent, stale, or not tied to the same live target. Tools should validate evidence against the live GUI before passing it to the agent.
\end{tcolorbox}

\subsection{Phase 2: Reasoning \& Planning Defects}
\label{subsec:P2_reasoning}
Before execution, P2 prepares the next interaction by constructing the model context, invoking the model, interpreting the selected action, and deciding whether to continue, stop, retry, or escalate. In all three patterns, the tool loop advances without a guard that checks whether the current state, payload, or environment permits the next step, producing 104 of 335 failures.

\medskip
\phantomsection
\label{subsubsec:infinite_loop}
\noindent\textbf{4. Progress Check Failure.}
Progress control depends on comparing consecutive tool outcomes and stopping, recovering, or changing strategy when repeated calls do not change observations, errors, or completion status. Without this guard, the loop treats another executable step as recovery and wastes steps or tokens on the same failed action. It affects 26 applications (65\%), averaging 3.08 cases each.

\begin{figure}
    \centering
    \includegraphics[width=0.9\linewidth]{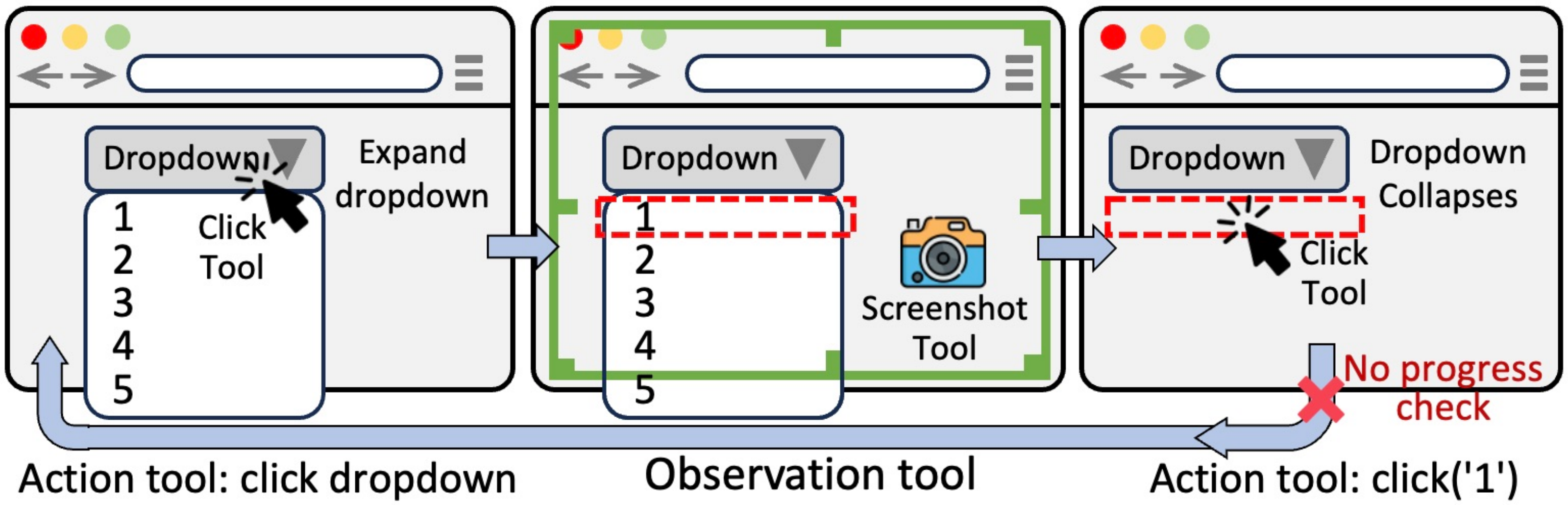}
    \vspace{-0in}
    \caption{Progress check failure after repeated unchanged observations in \app{computer\_use\_ootb}~\cite{computer_use_ootb2026}.}
    \label{fig:infinite_loop}
\end{figure}

Figure~\ref{fig:infinite_loop} shows the out-of-the-box GUI agent \app{computer\_use\_ootb}~\cite{computer_use_ootb2026} entering an ``open-screenshot-close'' cycle: the screenshot tool collapses the ``Numbers Covered'' dropdown before option selection, and without a dropdown-open check or repeated-action detector, the application reopens it indefinitely. The guard fails in different ways depending on what the loop is waiting for: the self-hosted virtual-computer agent \app{Bytebot}~\cite{bytebot} opens Firefox but never checks whether it has loaded, while the vision-free Windows GUI agent \app{Windows-Use}~\cite{windowsuse2024} keeps invoking \code{type\_tool} even after the target field stops responding to input.

\begin{figure}
    \centering
    \includegraphics[width=0.85\linewidth]{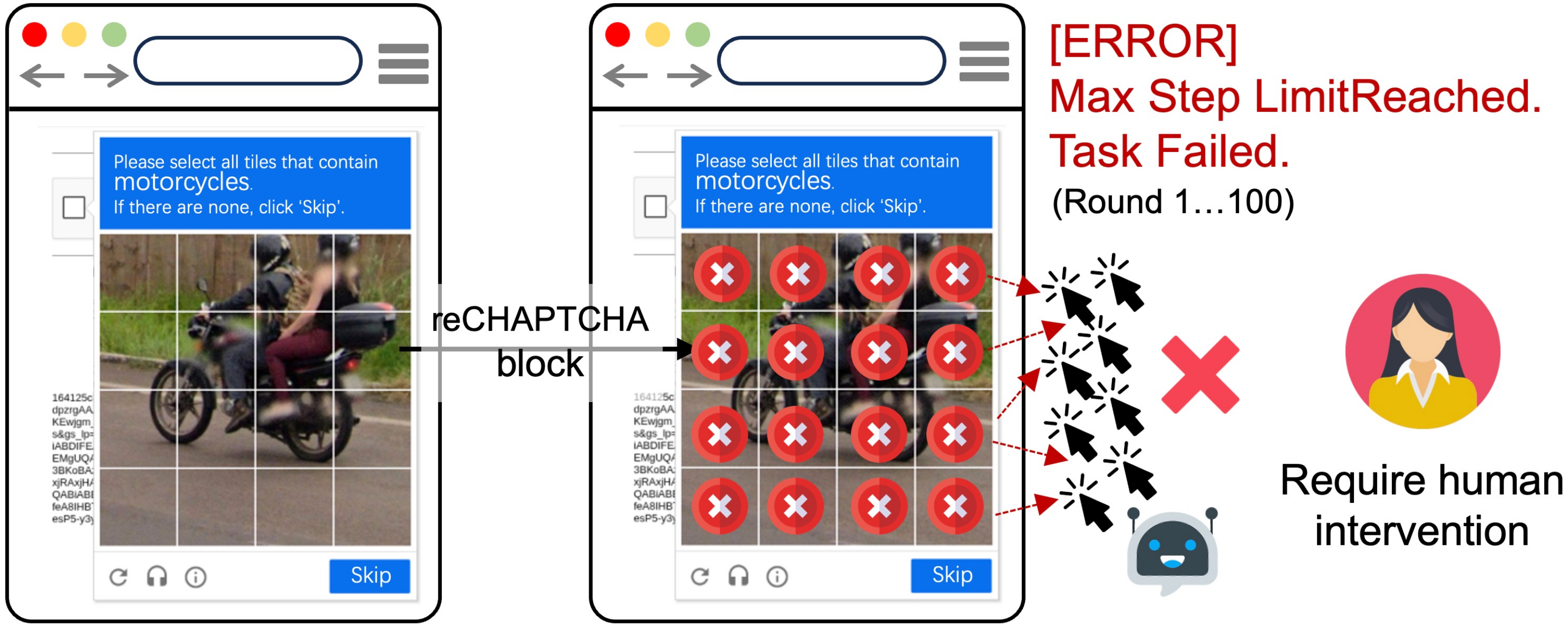}
    \vspace{-0in}
    \caption{Escalation boundary failure in \app{WebVoyager}~\cite{minorjerry_webvoyager_2024}.}
    \label{fig:permission}
\end{figure}

\medskip
\phantomsection
\label{subsubsec:permission_security}
\noindent\textbf{5. Escalation Boundary Failure.}
Escalation relies on observation tools distinguishing executable GUI states from security checkpoints that ordinary automation tools cannot pass~\cite{10.1145/2046707.2046779}, such as CAPTCHA, bot-detection, and human-verification barriers~\cite{jcp5030058,10.1145/2619091}. Returning these checkpoints as ordinary page states leaves action and navigation tools without a viable path forward. This rarer pattern affects 5\% of applications with a median of 1.5 cases.

Figure~\ref{fig:permission} shows the autonomous web-browsing agent \app{WebVoyager}~\cite{minorjerry_webvoyager_2024} triggering verification while searching for ``cakes'': its observation tools neither classify the CAPTCHA nor surface it as a stop signal, so redundant actions continue until the step budget is exhausted. A similar case appears in \app{LaVague}: when Google~\cite{google} blocks a video search with human verification, the browser tool returns a page that the application treats as actionable instead of security-blocked.

\medskip
\phantomsection
\label{subsubsec:token_limit_exceeded}
\noindent\textbf{6. Context Budget Overflow.}
Tool loops must bound observation, extraction, and history payloads before model invocation. Unpruned tool outputs can push the context window past model-interface limits before the next action is chosen. This overflow appears in 48\% of applications, with 21 cases in total.

The Playwright-based browser automation application \app{browser-use-typescript}~\cite{singh2025browser_use_typescript} illustrates the HTML-extraction variant: its browser tool invokes \code{page.content()}, converts the full page with \code{markdownify}, and can insert over 530,000 tokens against a 128,000-token limit. Other tool payloads fail for the same reason: the language-agent application \app{OpenAgents}~\cite{openagents} passes raw interface data from an observation tool, reaching 8,542 tokens against an 8,192-token model maximum, and the local automation application \app{autoMate}~\cite{autoMate2026} retains 20 high-resolution screenshots from a screenshot tool, totaling an estimated 255,503 tokens against a 199,000-token limit. In each case, the tool output is accepted without a size check, and the overflow is only detected after model invocation begins.

\begin{tcolorbox}[colback=black!5!white,colframe=black!80!white,boxsep=1pt,left=2pt,right=2pt,top=1pt,bottom=1pt,arc=0pt,outer arc=0pt,boxrule=0.7pt]
\emph{Finding-2:} P2 is the largest phase (31.0\%), where tool loops lack guards for payload size, non-progressing retries, or escalation boundaries. Tools should bound output size before model invocation, detect non-progressing states before retrying, and surface security barriers as stop signals.
\end{tcolorbox}

\subsection{Phase 3: Execution Defects}
\label{subsec:P3_execution}
P3 materializes a model-selected action as a concrete tool invocation and backend side effect. This phase depends on parsers, adapters, and executors that preserve the action across schema normalization, parameter binding, and readiness validation. In each case, tools accept, translate, or execute an action without checking whether it is valid, correctly parameterized, or safe to deliver to the current GUI or backend state, accounting for 82 of 335 failures.

\begin{figure}[t]
  \centering
  \captionsetup{font=small,skip=2pt}

  \begin{minipage}{0.82\linewidth}
\begin{lstlisting}[
  language=Python,
  escapechar=\%,
  basicstyle=\ttfamily\scriptsize,
  aboveskip=1pt,
  belowskip=1pt,
  lineskip=-1pt
]
raw = "x=331 y=79"
coords = {k: v for k, v in (i.split('=') for i in items)}
\end{lstlisting}
  \end{minipage}

  \caption{Coordinate parsing failure and fix in \app{pywinassistant}~\cite{pywinassistant2026}.}
  \label{fig:format_error_code}
\end{figure}

\medskip
\phantomsection
\label{subsubsec:format_schema_mismatch}
\noindent\textbf{7. Tool Schema Mismatch.}
Tool invocation begins with parsers and registries normalizing model output into registered actions with expected fields and types. Unhandled wrappers, schema variants, or rigid parsing rules turn recoverable action descriptions into unparseable or unregistered tool calls. This is the most widespread pattern, affecting 63\% of applications with a median of 2 cases each.

These mismatches break concrete parser and registry contracts, instead of merely changing surface syntax. One group fails at wrapper removal: in the desktop agent \app{screenAgent}~\cite{niuzaisheng_screenagent_2024}, Markdown-fenced JSON (\code{```json ... ```}) stops a parser that expects raw JSON. A second group reaches the tool interface but violates the registered action schema: \app{Windows-Use} generates names such as \code{functions.click\_tool} outside its registry. Figure~\ref{fig:format_error_code} illustrates the third group, parser rigidity: the Windows desktop GUI automation assistant \app{pywinassistant}~\cite{pywinassistant2026} crashes on \code{x=331 y=79} without commas, turning recoverable syntax variation into a hard failure.

\begin{figure}
    \centering
    \includegraphics[width=0.8\linewidth]{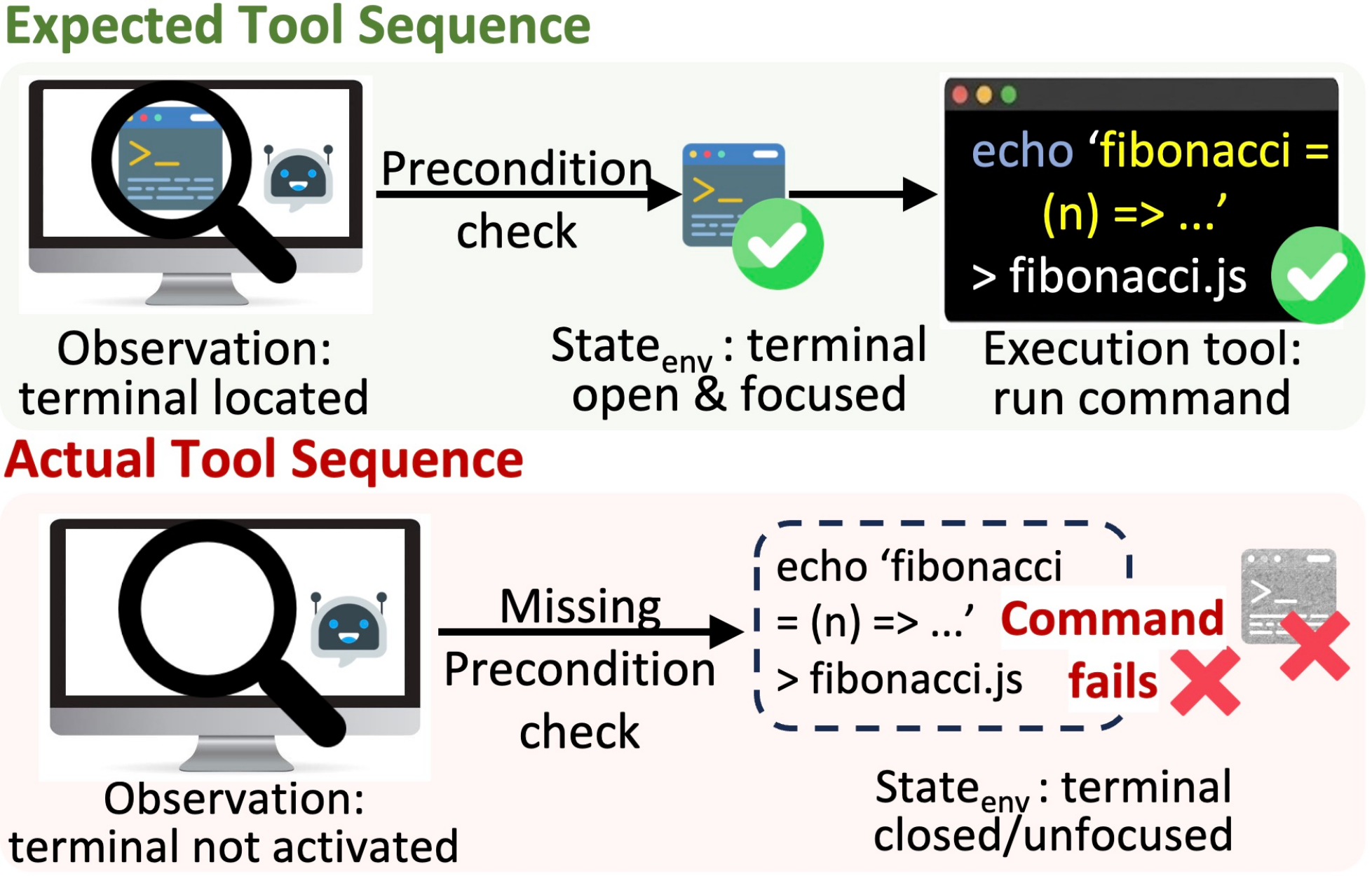}
    \vspace{-0in}
    \caption{Missing terminal-readiness precondition in \app{Surf}~\cite{e2b_surf2026} during command execution.}
    \label{fig:precondition_violation}
\end{figure}

\medskip
\phantomsection
\label{subsubsec:parameter_hallucination}
\noindent\textbf{8. Parameter Semantics Error.}
Tool adapters sit between a parseable call and executor invocation, where defaults are bound, arguments are rewritten, and placeholders are resolved. A call can satisfy the tool schema yet still execute the wrong operation if this adapter step changes what its parameters mean. The pattern affects 25\% of applications, averaging 1.70 cases per affected application.

Parameter semantics shift in the adapter before the call reaches the executor. In \app{autoMate}, the scroll adapter silently replaces a requested 200-pixel scroll with a 500-pixel system value, causing the action to overshoot the target. When placeholders go unresolved, the effect is more subtle: the natural-language web automation agent \app{actor-web-automation-agent}~\cite{apify2023web_automation_agent} sends a parseable CAPTCHA action whose value remains \code{<CAPTCHA\_TEXT>}, a marker that should instead become a structured execution error before the adapter forwards the call. Extra rewriting introduces a third variant: \app{pywinassistant}'s query adapter expands a concise search query into irrelevant generated content, changing the semantics of the action before it reaches the executor.

\medskip
\phantomsection
\label{subsubsec:precondition_violation}
\noindent\textbf{9. Missing Execution Preconditions.}
Execution tools sit at the boundary where a parsed call becomes a backend or GUI side effect. A call can be syntactically valid but still unsafe when the target resource or GUI state is not ready. Missing these readiness checks delivers shell, click, or type calls to missing resources or wrong GUI targets in 33\% of applications, with a median of 2 cases per application.

The failures arise from three sources: infrastructure unreadiness, planning order gaps, and input state staleness. Infrastructure unreadiness appears when backend resources are addressed before they exist: the computer-use application \app{cua}~\cite{cua2024} calls \code{get\_vm()} before provisioning or \code{run\_vm()} finishes, missing a VM-readiness gate before later execution tools address the backend resource. Planning order gaps arise when tools skip application setup: Figure~\ref{fig:precondition_violation} shows the virtual-desktop computer-use agent \app{Surf}~\cite{e2b_surf2026} issuing terminal commands before launching or focusing the terminal, so the execution tool targets a nonexistent window. Input state staleness exposes a subtler form of the same missing check: the Llama-4-powered computer-use agent \app{llama4-computer-use}~\cite{lee2024llama4} sends a social-media query before activating the browser, \app{AppAgent}'s text-entry tool types before activating a chat input field, and \app{gemini-browser}'s browser-input tool reuses a stale search bar between queries.

\begin{tcolorbox}[colback=black!5!white,colframe=black!80!white,boxsep=1pt,left=2pt,right=2pt,top=1pt,bottom=1pt,arc=0pt,outer arc=0pt,boxrule=0.7pt]
\emph{Finding-3:} Execution tools in P3 (24.5\%) accept malformed schemas, semantically invalid parameters, or actions without satisfied GUI preconditions. Tools should validate action schemas before invocation, reject unresolved parameter placeholders, and verify GUI and backend readiness before executing.
\end{tcolorbox}

\subsection{Phase 4: Environment Feedback Defects}
\label{subsec:P4_environment}
P4 carries the effect of the previous tool action into the next loop iteration. Feedback tools report completion, expose updated observations, and maintain resource handles for sessions, windows, APIs, and OS tools. Across both patterns, feedback tools report success or propagate state before the live GUI or resource has actually reached a stable, usable condition, producing 61 of 335 failures.

\begin{figure}
    \centering
    \includegraphics[width=\linewidth]{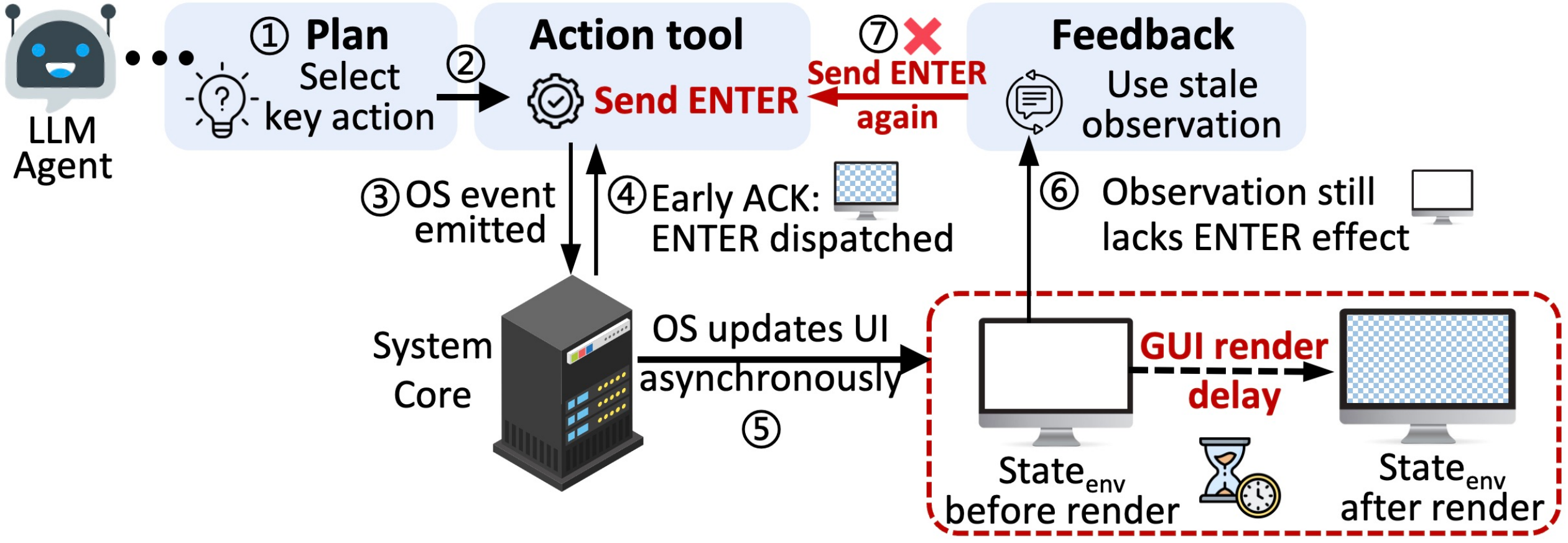}
    \vspace{-0in}
    \caption{Temporal feedback mismatch in \app{cua}~\cite{cua2024}: success is reported before the OS finishes rendering the UI update.}
    \label{fig:latency_mismatch}
\end{figure}

\medskip
\phantomsection
\label{subsubsec:latency_mismatch}
\noindent\textbf{10. Temporal Feedback Mismatch.}
Feedback tools report progress before the next planning cycle~\cite{10.1145/2489280.2489284,10.1145/2635868.2635920}. A tool that returns success or captures the page before the GUI reaches a stable, observable state leaves the next cycle planning from feedback that does not yet match $\text{State}_{\text{env}}$. We observe this pattern in 8 applications (20\%), averaging 1.38 cases each.

Figure~\ref{fig:latency_mismatch} shows \app{cua} returning success for key operations before the OS has finished rendering the UI update: the action tool dispatches the key event and receives an early acknowledgment, but the GUI render delay means the subsequent observation still reflects the pre-action state. The timing contract also governs text input and DOM mutation: \app{Clevrr-Computer}~\cite{clevrr-computer} drops spaces during fast programmatic text entry, while \app{UI-TARS-desktop}'s \code{showWaterFlow} injects elements before \code{document.head} or \code{document.body} exists. In each case, the missing check is state stability before the next observation is captured.

\medskip
\phantomsection
\label{subsubsec:unexpected_interruption}
\noindent\textbf{11. Lifecycle Propagation Failure.}
Feedback tools carry lifecycle state for sessions, windows, pages, OS tools, and APIs. When disappearance, crashes, or incompatible return data are not propagated as recoverable structured errors, the agent keeps using stale handles or converts backend failure into ordinary continuation. This pattern affects 23 applications (58\%), averaging 2.17 cases each.

When a resource disappears without a structured error, the agent continues against a stale handle. \app{Surf} illustrates this path: during a flight-booking task, the system tries to extend the E2B sandbox only after it has already been destroyed, producing \code{SandboxError: 404: Sandbox not found}; at that point, the desktop connection is gone and later tool calls cannot recover the task. Related cases treat a timed-out or black-screen sandbox as a recoverable UI state, so the agent keeps taking screenshots, refreshing, clicking, or sending keys against a dead environment.

\begin{figure}
    \centering
    \includegraphics[width=\linewidth]{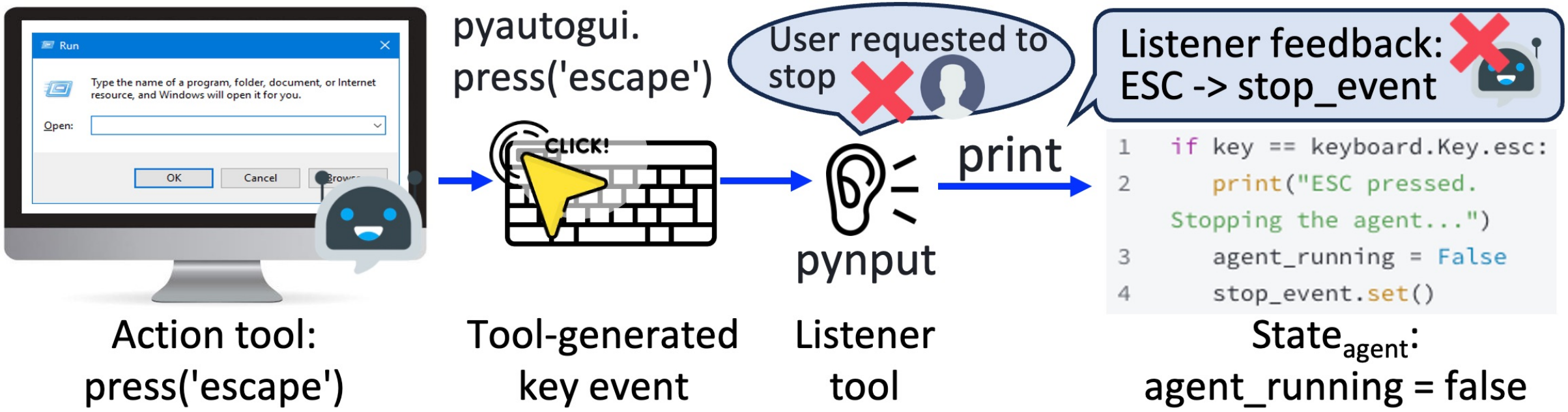}
    \vspace{-0in}
    \caption{Lifecycle propagation failure in \app{desktop-agent}~\cite{koroy2024desktop_agent} triggered by a misinterpreted \code{ESC} signal.}
    \label{fig:unexpected_interruption}
\end{figure}

Lifecycle state can also be misrouted instead of lost. Figure~\ref{fig:unexpected_interruption} shows the desktop automation agent \app{desktop-agent}~\cite{koroy2024desktop_agent} issuing \code{pyautogui.press('escape')} to close a failed Run dialog. The \code{pynput} listener tool intercepts this tool-generated key event and treats it as a user-stop signal, setting \code{agent\_running = false} and terminating the run. The action tool had no way to mark its key event as internal, so the listener tool propagated it as a lifecycle stop without provenance.

\begin{tcolorbox}[colback=black!5!white,colframe=black!80!white,boxsep=1pt,left=2pt,right=2pt,top=1pt,bottom=1pt,arc=0pt,outer arc=0pt,boxrule=0.7pt]
\emph{Finding-4:} Although P4 is the smallest phase (18.2\%), its failures are easy to overlook: feedback tools must verify GUI stability before reporting success and propagate resource lifecycle events as structured errors instead of ordinary continuation signals.
\end{tcolorbox}

\section{Solution: \tool}
\label{sec:solutions}

\begin{figure}[t]
    \centering
    \includegraphics[width=\linewidth]{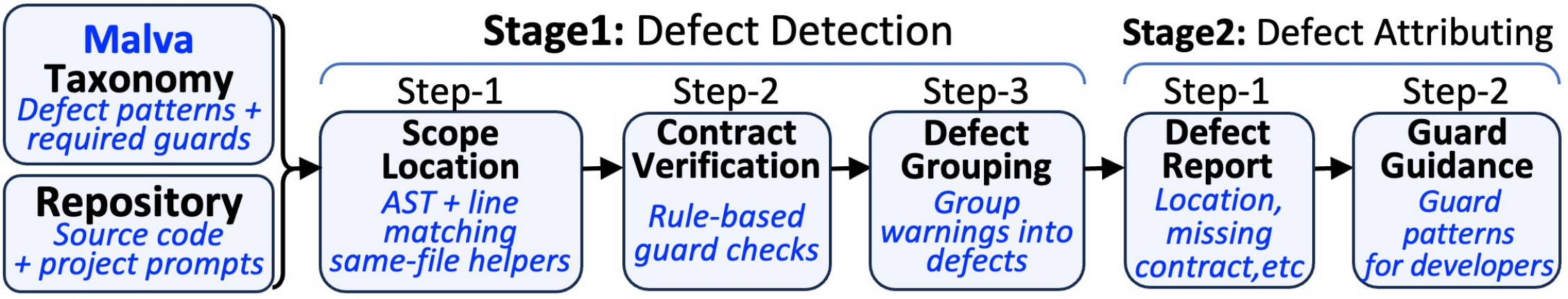}
    \vspace{-0.1in}
    \caption{Two-stage workflow of \tool.}
    \label{fig:malvaguard_workflow}
\end{figure}

Section~\ref{sec:defect} shows that many runtime interaction failures stem from code-level anti-patterns: tool-layer operations are invoked, retried, or accepted without guards that align the agent state with the live environment. Findings~1--4 confirm that these missing guards appear systematically across all four phases. To detect them before execution, we implement \tool, a pre-deployment static checker~\cite{10.1145/3188720,7194609}. As shown in Figure~\ref{fig:malvaguard_workflow}, \tool encodes the taxonomy as pattern-based static analysis rules.

\subsection{Stage 1: Defect Detection}

\textbf{Scope location.}
\tool first identifies the scope to check. We define an \emph{interaction region} as the local code or prompt span that contains a tool-environment interaction: in source code, it is the enclosing function or method; in project prompts or files without supported syntax trees, it is the matched line together with nearby lines. \tool searches source-code abstract syntax trees for operations that invoke models, read observations, handle actions, or update backend state, and identifies the corresponding interaction region for each match.
\tool also checks helper calls (\eg, parsing and execution) within the same code file. If the helper contains tool execution, parsing, or verification logic, \tool checks that logic as part of the caller's region.

\textbf{Contract verification.} \tool applies rule-based checks derived from the taxonomy. Each rule pairs a trigger pattern for a tool-environment interaction with a required guard: code or prompt evidence that bounds observations, constrains loop progress, validates tool calls, or verifies feedback before $\text{State}_{\text{agent}}$ is updated from tool results. \tool searches for the guard within the interaction region; if absent, \tool creates a missing-contract warning with the source location and missing guard.

\textbf{Defect grouping.} \tool generates a warning for each missing-contract location identified in the previous step. Within the same tool helper, adapter, or agent module, a single missing contract can trigger multiple warnings: the tool call, action parsing, retry logic, and feedback handling may all be flagged by the same checker rule because they share the same guard gap. \tool groups these co-located warnings into one defect only when they share the same checker rule and missing-contract type, recording one representative source line and keeping the remaining warnings as supporting evidence. Warnings from different checker rules are reported as separate defects even when they appear in the same component.

\subsection{Stage 2: Defect Attributing}

For each defect produced by Stage 1, \tool generates a structured report and guard suggestion for developers.

\textbf{Defect report.} For each defect, \tool records its location (file path, line number, and enclosing tool helper, adapter, or agent module), the missing contract (\eg, observation bound, action validation, progress check, or feedback verification), the corresponding taxonomy defect pattern (\eg, Context Budget Overflow, Tool Schema Mismatch), and the warning locations merged as supporting evidence.

\textbf{Guard guidance.} \tool suggests a guard pattern as a repair solution for each defect, covering observation bounds, loop progress, action validation, and feedback verification across all four phases. The guidance identifies which class of check to add at the reported location, instead of providing an automated fix.

\subsection{Implementation}
\tool supports Python, JavaScript, and TypeScript. It uses Python \code{ast}~\cite{python_ast} and Tree-sitter frontends for source code~\cite{treesitter2026} and line-based prompt matching. All experiments run on a cloud instance with 10 CPU cores, 100 GB system memory, 100 GB shared memory, an NVIDIA GeForce RTX 4090 GPU (48 GB VRAM, driver 550.163.01, CUDA 12.4), and Samsung NVMe SSD storage.

\subsection{Evaluation}
We evaluate \tool on a held-out corpus of 20 applications outside \defectlib, with no overlap with the original 40, following the same selection criteria as Section~\ref{sec:methodology}. We manually build a defect oracle from issue, commit, and execution evidence using the same validation process as Section~\ref{sec:methodology}, yielding 112 real defects. \tool flags 497 missing-contract locations and groups them into 150 defects. Of these, 86 match oracle defects and 64 do not, yielding 57.3\% precision. \tool misses 26 oracle defects, yielding 76.8\% recall.

\begin{figure}
    \centering
    \includegraphics[width=\linewidth]{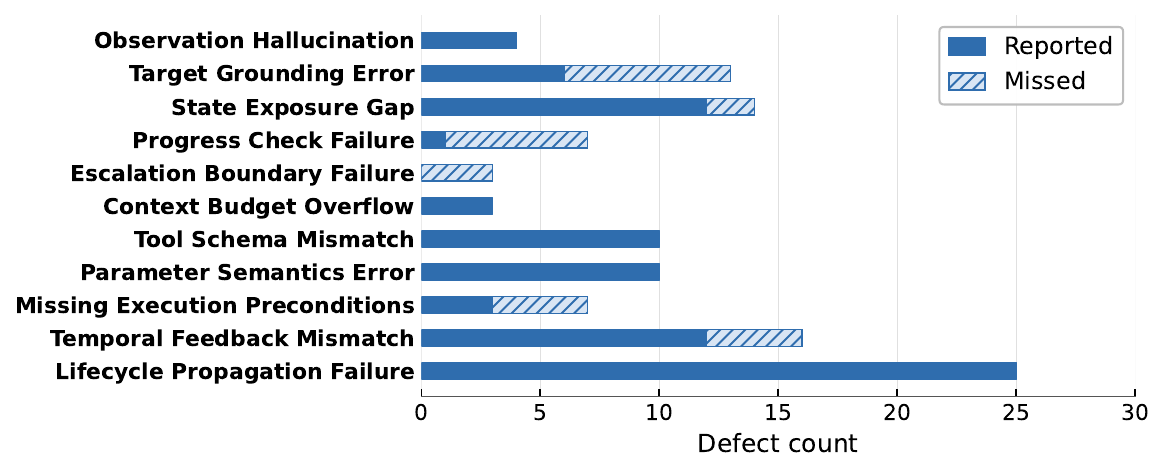}
    \vspace{-0.1in}
    \caption{\tool's reported and missed real defects by defect pattern on the additional corpus.}
    \label{fig:malvaguard_coverage}
\end{figure}

Figure~\ref{fig:malvaguard_coverage} breaks down the reported and missed defects by pattern. \tool achieves full recall on Tool Schema Mismatch, Parameter Semantics Error, Context Budget Overflow, and Lifecycle Propagation Failure, where the missing guard is co-located with the triggering operation in the same local scope. Recall is lower for Progress Check Failure, Missing Execution Preconditions, and Target Grounding Error because their missing guards span multiple components or depend on cross-file conventions not visible in local source. Escalation Boundary Failure has zero reported cases because its instances are runtime security barriers, not guard omissions visible in local code.

Both error types stem from the limits of local static analysis. False positives arise when a guard is hidden in framework wrappers, inherited callbacks, or runtime checks outside the local scope. They also arise when guard semantics span exception-handling and fallback paths. False negatives arise when defects depend on cross-file reasoning or framework-specific adapter semantics not visible in local code. These results reveal a broader code-quality risk: many applications still lack explicit code-level guards around tool use, leaving agent-environment interactions weakly protected regardless of whether a static checker is applied.

\subsection{Case Study}
We examine \app{skyvern}~\cite{skyvern2026}, a web-browser agent, to illustrate \tool's reporting and limits. Its browser helper invokes \code{page.act(...)} and immediately returns \code{ActResult(..., completed=True)} without re-observing the page or verifying the action outcome. \tool locates the helper as the interaction region, finds no such guard, and reports a Temporal Feedback Mismatch defect, citing the action call as evidence and suggesting a postcondition guard on the observable GUI state. This is a confirmed true positive. 

The same application, however, also produces a false positive that illustrates the limits of local static analysis: \tool flags \code{await asyncio.sleep(0.3)} after an \code{Escape} keypress clearing a possible overlay before a selector retry under the same pattern. Manual validation rejects this because the sleep belongs to a recovery path that falls back to the LLM planning loop on failure, showing how local static analysis mistakes a recovery wrapper for a missing guard when the guard semantics span exception handling and fallback logic.

\section{Threats to Validity}

\textbf{Internal Validity.}
Manual labels may contain bias or miss rare patterns despite multi-annotator confirmation. GitHub keyword filtering may miss unreported failures and issues described in non-technical terms. Reproduction may omit environment-specific factors that affect whether a failure manifests.
\tool is a static checker and relies on local source and prompt evidence. It may miss contracts that become visible only at runtime or outside the local source scope.

\textbf{External Validity.}
Our findings may not generalize to all agentic GUI software. Windows/Linux behavior may differ on other platforms, and the common-use scenarios in Section~\ref{sec:methodology} may miss specialized workflows in enterprise, accessibility, or hardware-control settings. Our open-source focus enables source-level confirmation from issues, commits, and execution logs, but may miss closed-source constraints or mitigations.

\section{Related Work}

\sloppy

\textbf{Empirical Studies and Defect Analysis.}
Prior works have studied defects across various software systems. One line of works studied defects in traditional software, including large-scale studies on code quality~\cite{large-scale_study}, learning-based name bug detection~\cite{DeepBugs}, and defect prediction surveys~\cite{Literature}. Another line of works studied ML-enabled software, building taxonomies of faults in deep learning systems~\cite{10.1145/3377811.3380395}, examining whether ML cloud APIs are used correctly~\cite{ML_API}, preventing their integration failures at runtime~\cite{Run-Time_Prevention}, and automatically testing and fixing ML software when misbehavior is exposed~\cite{keeper}. The same theme extends to surveying ML testing landscapes~\cite{9000651}. Recent work further studies LLM and RAG integration defects in open-source software~\cite{llm_study}, bugs in LLM-generated code~\cite{10.1007/s10664-025-10614-4}, and failure attribution in multi-agent systems to automatically identify which agent and step are responsible for task failures~\cite{zhang2025which}. 
These works focus on code, model pipelines, API usage, or generated artifacts, but do not address failures that arise at the boundary between an LLM agent and a live GUI environment. This paper fills this gap by studying defects at this tool-mediated boundary and building \defectlib and \tool to support detection.

\textbf{LLM Agents and GUI Automation.}
LLM agents combine planning, memory, and tool use~\cite{survey,DBLP:journals/chinaf/XiCGHDHZWJZZFWXZWJZLYDW25,yao2023react,10.1145/3586183.3606763,schick2023toolformer,huang2025evaluating}. Recent systems and benchmarks evaluate web navigation~\cite{zhou2024webarena,deng2023mindweb}, code generation~\cite{jimenez2024swebench}, desktop/OS control~\cite{OSWorld,chezelles2025the}, and mobile automation~\cite{rawles2023androidinthewild,10.24963/ijcai.2024/711}, while frameworks such as OpenClaw study GUI-agent architectures~\cite{openclaw2026}. Several studies further analyze capability and benchmark-level weaknesses such as localization bias, visual grounding, and unintended consequences in human-agent browsing~\cite{zhang2025large,tao-etal-2025-understanding,wu2026guiactor,10.1145/3637528.3671650,OSWorld,rawles2025androidworld,10.1145/3772318.3790696}.
These works focus on whether an agent eventually completes a task, instead of how the tools that mediate between the agent and the environment fail to do so reliably. Our scope is the tool layer itself: how and why it fails to keep agent and environment states aligned.

\textbf{GUI Testing and Test Automation.}
Prior works have studied GUI testing through several approaches. One line of works explored target-application defects through record-and-replay or script-based automation~\cite{10.1145/1622176.1622213,10.5555/2486788.2486799,selenium2026,appium2026} and random or model-based exploration covering stochastic input generation, state-machine-guided traversal, and use-case combinations~\cite{monkey,gremlins,machiry2013dynodroid,patel2018effectiveness,su2017guided,guitar,li2017droidbot,yang2013grey,wang2020combodroid,mesbah2008crawling}. Another line of works applied AI/LLM-assisted methods guided by semantics, functionality, or natural language~\cite{10.1109/ICSE-Companion58688.2023.00024,10.1145/3468264.3468554,10.1145/2950290.2983958,zheng2021automatic,chang2023reinforcement,liu2024make,liu2025seeing}, including deep learning-based black-box approaches to Android app testing~\cite{8952324}.
All of these works treat automation code as testing infrastructure and the target application as the subject of study. In agentic GUI software, however, the tool layer is part of the deployed system itself and participates in ordinary user tasks. This paper shifts the defect boundary to the tool layer itself, treating observation, action, and feedback tools as the primary subject of analysis instead of the vehicle for testing other software.

\section{Conclusion}

Agentic GUI software uses observation, action, and feedback tools to let LLM agents operate live interfaces. This paper analyzes 335 confirmed failures from 2,000 issues across 40 applications, identifying 11 defect patterns rooted in state representation, transition, and synchronization problems that degrade functionality, efficiency, and reliability across 90\% of the studied applications. We release \defectlib, a defect library of confirmed failures with their root causes and impacts, and \tool, a pre-deployment static contract checker that reaches 57.3\% precision and 76.8\% recall on an additional 20-application corpus. We hope this work can serve as a starting point for improving the reliability of the tool layer in agentic GUI software.

\section{Data Availability}

The artifact is publicly available at \url{https://doi.org/10.6084/m9.figshare.32838281}, including the studied applications, the defect library \defectlib, the source code of \tool, detailed experimental results, and a usage guide.

\newpage
\bibliographystyle{IEEEtran}
\bibliography{citation.bib}

\end{document}